\documentclass[aps,prl,reprint,superscriptaddress,amsmath,amssymb,floatfix]{revtex4-2}
\usepackage{amsmath}
\usepackage{amssymb}
\usepackage[utf8]{inputenc}
\usepackage{array}
\usepackage{multirow}
\usepackage{esint}
\usepackage{dcolumn}
\usepackage{bm}
\usepackage{bbm}
\usepackage[breaklinks,colorlinks,
linkcolor=blue,citecolor=blue,urlcolor=blue]{hyperref}
\usepackage{babel}
\usepackage{titlesec}
\usepackage{graphicx}
\usepackage{mathtools}
\usepackage{soul}
\usepackage[dvipsnames]{xcolor}
\usepackage{tikz}
\usepackage{tcolorbox}
\usepackage{physics}
\usepackage{orcidlink}

\begin{document}

\preprint{APS/123-QED}

\title{
%Magnetostriction Crossover via \textit{in-situ} Images of Quasi-2D Strongly Dipolar Bose Gases\\
\textit{In-situ} Observation of Magnetostriction Crossover in a Strongly Dipolar Two-Dimensional Bose Gas\\
%Magnetostriction Crossover via \textit{in-situ} Imaging of two-dimensional Strongly Dipolar Bose Gases
}% Force line breaks with \\

\author{Yifei He\orcidlink{0000-0003-0713-6376}}
\thanks{These authors contributed equally.}
\email{Present address: Department of Physics, The University of Hong Kong, Pok Fu Lam, Hong Kong, China}
\affiliation{Department of Physics, The Hong Kong University of Science and Technology, Clear Water Bay, Kowloon, Hong Kong, China}

\author{Xin-Yuan Gao\orcidlink{0000-0003-0608-9029}}
\thanks{These authors contributed equally.}
\affiliation{Department of Physics, The Chinese University of Hong Kong, Shatin, Hong Kong, China}

\author{Haoting Zhen\orcidlink{0009-0001-9463-8480}}
\affiliation{Department of Physics, The Hong Kong University of Science and Technology, Clear Water Bay, Kowloon, Hong Kong, China}
\affiliation{Department of Physics and Astronomy, Rice University, Houston, TX, USA}

\author{Mithilesh K. Parit\orcidlink{0000-0003-4375-7014}}
\affiliation{Department of Physics, The Hong Kong University of Science and Technology, Clear Water Bay, Kowloon, Hong Kong, China}

\author{Yangqian Yan\orcidlink{0000-0002-3237-5945}}
\email{yqyan@cuhk.edu.hk}
\affiliation{Department of Physics, The Chinese University of Hong Kong, Shatin, Hong Kong, China}
\affiliation{State Key Laboratory of Quantum Information Technologies and Materials, The Chinese University of Hong Kong, Hong Kong SAR, China}
\affiliation{
The Chinese University of Hong Kong Shenzhen Research Institute, Shenzhen, China
}%

\author{Gyu-Boong Jo\orcidlink{0000-0002-0500-3572}}
\email{gbjo@rice.edu}
\affiliation{Department of Physics and Astronomy, Rice University, Houston, TX, USA}
\affiliation{Smalley-Curl Institute, Rice University, Houston, TX, USA}
\affiliation{Department of Physics, The Hong Kong University of Science and Technology, Clear Water Bay, Kowloon, Hong Kong, China}

\begin{abstract}
Magnetostriction, the anisotropic spatial deformation, is a hallmark of dipolar gases with strong long-range interactions, yet it poses a challenge for {\it in-situ} characterization. Here, we observe a magnetostriction crossover from the strongly anisotropic superfluid phase to the nearly isotropic normal phase using {\it in-situ} imaging of quasi-two-dimensional $^{166}$Er gases. Then, we develop a quasi-2D Hartree-Fock-mean-field framework that provides a robust tool for interaction-aware thermometry, enabling the determination of temperature and chemical potential across all dipole orientations from a single fit. We further demonstrate that the low-density wings effectively obey a local-density equation of state. Finally, we reveals the crossover from the isotropic thermal wings to the anisotropic coherent core in a single \textit{in-situ} image, providing a pathway for future accurate studies of strongly dipolar superfluidity and thermodynamics in 2D.
\end{abstract}

\maketitle

\textit{Introduction.}---Local probing in atomic samples has opened new opportunities to investigate many-body physics beyond the global measurements~\cite{ku2012revealing,bakr2009quantum,gemelke2009situb,bakr2010probing,sherson2010single,mazurenko2017cold}. For example, \textit{in-situ} thermometry has been widely applied to atoms with contact interactions, especially in two-dimensional~(2D) systems~\cite{tung2010observation,hung2011observation, yefsah2011exploring}. Using the local-density approximation (LDA), one can extract the equation of state (EOS) across the superfluid-to-normal transition and determine the critical point~\cite{hung2011observation,yefsah2011exploring,zhang2012observation} from trapped systems, or characterize the Mott insulator~\cite{gemelke2009situb,bakr2010probing,sherson2010single}.  While these techniques are well-established for contact-interacting gases, their application to dipolar atomic systems, such as magnetic atoms~\cite{lahaye2009physics,chomaz2022dipolar,griesmaier2005bose,lu2011strongly,aikawa2012bose}, polar molecules~\cite{ni2008high,takekoshi2014ultracold,park2015ultracold,bigagli2024observation,shi2025boseeinstein}, and Rydberg-dressed systems~\cite{schauss2012observation,guardado2021quench,weckesser2025realization}, remains a significant challenge due to their long-range, anisotropic interactions that fundamentally distinguish them from short-range interacting systems.

This challenge arises from the spatial deformation, also known as magnetostriction~\cite{stuhler2005observation}, induced by the long-range dipole-dipole interaction (DDI) in dipolar gases. While standard thermometry holds for dipoles perpendicular to the 2D plane~\cite{ticknor2012finite}, it remains an open question how the LDA-based analysis can be performed when the magnetostriction induced by large tilt angles and strong DDI becomes dominant. Just as explicitly accounting for anisotropic DDI in dipolar thermal gases near the critical temperature of Bose-Einstein condensation (BEC) is necessary for accurate time-of-flight (TOF) thermometry~\cite{tang2016anisotropic}, understanding the finite-temperature \textit{in-situ} magnetostriction is crucial for studying thermodynamics in 2D. Earlier work demonstrates the EOS can be measured for a 2D dipolar gases when local interactions are dominant~\cite{he2025exploring}. However, its behavior in the strongly dipolar regime remains largely unexplored~\cite{he2025review}.

In this Letter, we address these gaps through \textit{in-situ} imaging of strongly dipolar quasi-2D $^{166}$Er gases with a tunable dipole orientation. We focus on thermal clouds with temperature close to the superfluid transition, where interaction effects are essential to account for thermodynamics~\cite{hung2011observation,he2025exploring}. We observe that while the superfluid core exhibits pronounced magnetostriction when dipoles are tilted into the plane, the thermal cloud remains magnetostriction-free. To explain this,  we develop a quasi-2D Hartree-Fock-mean-field (HFMF) theory with DDI that quantitatively describes normal-phase density profiles across all dipole angles. This provides a reliable thermometry framework: fitting the measured profiles with our formalism yields consistent results. We demonstrate that the low-density wings obey a local-density EOS, validating LDA-based thermodynamic extraction in the strongly dipolar regime. Finally, by imaging density contours near the superfluid transition, in a single \textit{in-situ} image, we reveal a smooth magnetostriction crossover from the isotropic thermal periphery to the anisotropic coherent core, offering a new route for future studies of superfluidity and beyond-mean-field physics in dipolar systems.

Our work enables accurate determination of temperature and chemical potential in dipolar atoms and molecules, providing essential tools for precisely characterizing their exotic phases, including magnetostriction~\cite{wenzel2018anisotropic,wenzel2017striped}, crystalline order~\cite{kadau2016observing,schmitt2016self,wenzel2017striped}, supersolidity~\cite{guo2019low,norcia2021two,chomaz2018observation,tanzi2019observation,bottcher2019transient,chomaz2019long,zhen2025breaking,he2025observation}, and quantized vortices~\cite{klaus2022observation,casotti2024observation}. Especially, it opens new routes for exploring 2D dipolar systems with thermodynamics and excitations~\cite{su2023dipolar,he2025exploring,zhen2025breaking,he2025observation}.

\begin{figure}[t!]
\includegraphics[width=1\columnwidth]{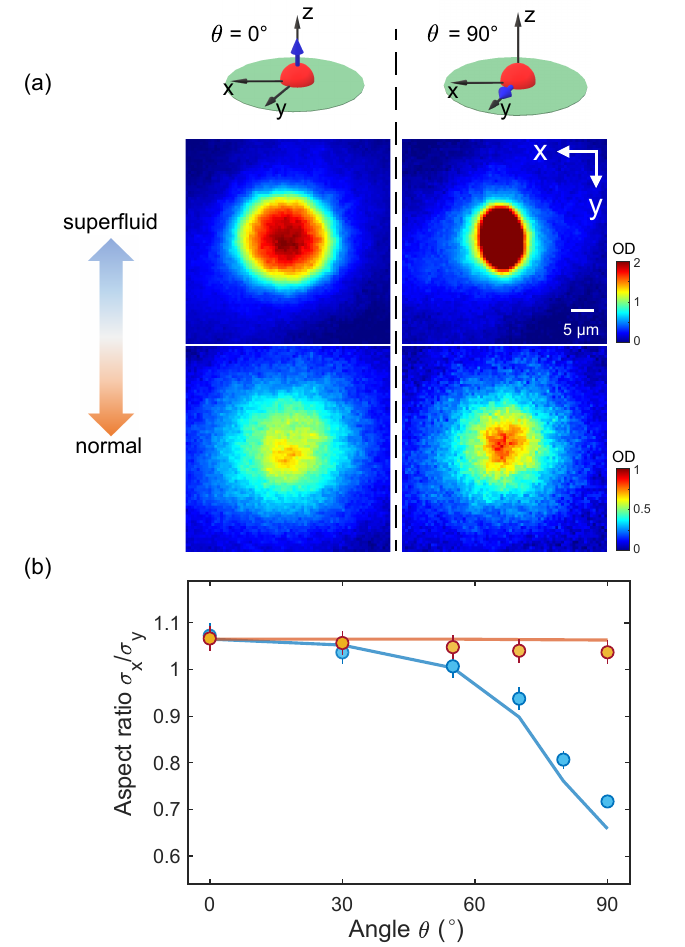}% Here is how to import EPS art
\caption{\textbf{Magnetostriction in superfluid and normal phases of quasi-2D dipolar gases.} (a)~\textit{In-situ} images of quasi-2D clouds of $^{166}$Er atoms with $\theta=0^\circ$~(left) and $\theta=90^\circ$~(right). The upper row shows samples with a superfluid core; the lower row shows purely thermal gases. 
%The optical density of normal gases has been doubled for clarity. 
(b)~The aspect ratio of superfluid core~(blue) and purely thermal cloud~(orange) as a function of $\theta$. The aspect ratio of superfluid cores is obtained from the Thomas-Fermi radii of the central parabolic part extracted via a bimodal fit. The aspect ratio of thermal clouds is extracted from a 2D Gaussian fit. Error bars reflect the 95\% confidence interval of the fits. The blue (orange) solid line is computed using the dipolar GPE (HFMF theory).}
\label{insitu} 
\end{figure}

\textit{Experimental Procedure.---} We prepared ultracold $^{166}$Er atoms in a quasi-2D harmonic trap with trap frequency $(\omega_x,\omega_y,\omega_z)=2\pi\times(20.0(1),21.3(2), 983(5))~\text{Hz}$, following our previous sequence~\cite{he2025exploring}. The interaction $U$ between erbium atoms reads:
\begin{equation}
    U(\mathbf{r})= g\delta(\mathbf{r})+\frac{3g_d}{4\pi}\dfrac{1-3\left(z\cos\theta+y\sin\theta\right)^{2}/r^2}{r^3},
\end{equation}
where $\mathbf{r}$ is the relative position vector, $r=\abs{\mathbf{r}}$ its magnitude, and $\theta$ denotes the dipole tilt angle with respect to the $z$ axis.
The experiments are carried out under a magnetic field of 900 mG with varying dipolar orientations $\theta$ and an $s$-wave scattering length of $a_s={m g}/{4 \pi \hbar^2}=67(3)~a_0$. The magnetic moment $\mu_m$ of erbium atoms is $7~\mu_B$, which gives rise to a dipolar length $a_{dd}={m g_d}/{(4 \pi \hbar^2)}=m\mu_0\mu_m^2/(12\pi\hbar^2)=65.5 a_0$ and relative dipolar strength $\epsilon_{dd}=a_{dd}/a_s=0.98(4)$. 
The uncertainty arises from the calibration of the background $s$-wave scattering length~\cite{patscheider2022determination}. After setting the magnetic field orientation to the target angle, we wait for the cloud to reach equilibrium, then take \textit{in-situ} absorption images. Superfluid samples are obtained at temperatures around $43~\text{nK}$ with $T/T_c\approx0.35$, while samples in the normal phase are obtained by lowering the atom number and increasing the temperature to around $70~\text{nK}$, resulting in $T/T_c\approx1.1$, where $T_c=\sqrt{6N_0}\hbar\sqrt{\omega_x\omega_y}/\pi k_B$ is the critical temperature of the ideal BEC in the 2D harmonic trap, and $N_0$ is atom number in axial ground state. 

Although the thermal wavelength is comparable to the axial confinement length, transverse condensation ensures the majority of atoms occupy the axial ground state~\cite{chomaz2015emergence}. This justifies a quasi-2D treatment of the interaction, which we decompose into local $U_\text{eff}^\text{l}$ and nonlocal $U_\text{eff}^\text{nl}(\mathbf{k})$ contributions~\cite{fischer2006stability,nath2009phonon,ticknor2011anisotropic}:
\begin{equation}
\begin{aligned}
    U_\text{eff}(\mathbf{k})=&U_\text{eff}^\text{l}+U_\text{eff}^\text{nl}(\mathbf{k}),\;U_\text{eff}^\text{l}=g_\text{eff}\equiv\frac{g+(3\cos^2\theta-1) g_d}{\sqrt{2\pi}l_z},\\
    U_\text{eff}^\text{nl}(\mathbf{k})=&\frac{3g_d}{2} k e^{\frac{l_z^2k^2}{2}}\operatorname{erfc}\left(\frac{l_z k}{\sqrt{2}}\right)\left[\frac{k_y^2\sin^2\theta}{k^2}-\cos^2\theta\right],
\end{aligned}
\end{equation}
where $\mathbf{k}$ denotes momentum in the $x-y$ plane, and $l_z=\sqrt{\hbar/m \omega_z}$ is the axial harmonic oscillator length. 

Figure~\ref{insitu}(a) shows exemplary \textit{in-situ} images. When the dipoles are perpendicular to the 2D plane (i.e., $\theta=0$), the interaction remains isotropic, and the shape of the cloud closely follows the trap. When the dipoles are tilted into the $x-y$ plane (i.e., $\theta=90^\circ$), the isotropic local $g_\text{eff}=g(1-\epsilon_{dd})/(\sqrt{2\pi}l_z)$ is almost canceled because $\epsilon_{dd}\simeq 1$, and the anisotropic nonlocal term becomes dominant in the interaction. Due to the highly anisotropic interaction, we observe a significant magnetostriction effect in the superfluid sample at $\theta=90^\circ$ (see the upper right panel of Fig.~\ref{insitu}(a)). In the normal phase, however, we observe an absence of magnetostriction, as shown in the lower panel of Fig.~\ref{insitu}(a). When tuning $\theta$ from $0^\circ$ to $90^\circ$ adiabatically in normal gases, the clouds shrink almost isotropically, aligning with the decrease in isotropic $g_\text{eff}$. To quantify this difference, we measure the aspect ratios at different $\theta$ in both superfluid and purely thermal samples, as shown in Fig.~\ref{insitu}(b). 
The superfluid aspect ratio, derived from the Thomas-Fermi radius via bimodal fitting, changes significantly above $55^\circ$, consistent with 2D dipolar Gross-Pitaevskii equation (GPE) predictions~\cite{cai2010mean}. In contrast, the thermal cloud aspect ratio, extracted via 2D Gaussian fitting, remains angle-independent and magnetostriction-free.

\textit{Theoretical Description of Normal Phase.---}
To elucidate the magnetostriction-free character of the thermal cloud, we apply the quasi-2D HFMF theory~\cite{kadanoff2018quantum,fetter2012quantum}. Because the axial confinement is much larger than any other relevant energy scale, the single-particle spectrum can be factorized into axial harmonic levels and transverse plane-wave states,
$
\varepsilon^{(0)}_{j,\mathbf k}= \hbar^{2}k^{2}/2m -\mu + j\hbar\omega_{z}
\;(j=0,1,\dots)
$,
while the atoms experience an in-plane harmonic potential $V(\mathbf r)=(m\omega_{x}^{2}r_x^{2}+m\omega_{y}^{2}r_y^{2})/2$.
In the experimental temperature regime, due to the low density, we treat the axial degrees of freedom as ideal Bose excitations, i.e., inter-axial-band interactions are neglected (truncating at \(j=0\) falls back to the mean-field theory).
Gradient expanding the Wigner transformed lesser Green's function up to the second order~(see End Matter), we define the phase space density for the $j$ band of the system to be
\begin{equation}
\begin{aligned}
f_j(\mathbf r,\mathbf k) = f_j^{\text{TF}}(\mathbf r,\mathbf k)+f_j^{\text{vW}}(\mathbf r,\mathbf k)+\cdots,\;f_j^{\text{TF}}=g_B({E}_j),
\end{aligned}
\label{phase_space_distribution}
\end{equation}
where the phase-space energy ${E}_j(\mathbf{r},\mathbf{k})$ is ${E}_j=\varepsilon^{(0)}_{j, \mathbf k}
+V(\mathbf r)
+\delta_{j,0}\hbar\Sigma^{\star}(\mathbf r,\mathbf k)$, and $g_{B}(E)=1/(e^{E/k_BT}-1)$ is the Bose-Einstein distribution function (see End Matter for the definition of the von-Weizs\"acker term $f_j^{\text{vW}}$).
The Hartree-Fock proper self-energy $\hbar\Sigma^{\star}(\mathbf r,\mathbf k)$ also consists of local and non-local parts $\Sigma^{\star}_\text{l}(\mathbf r) +
\Sigma^{\star}_\text{nl}(\mathbf r, \mathbf{k})$, with local contribution $\hbar\Sigma^{\star}_\text{l}=2 g_\text{eff} n_0(\mathbf{r})$ and non-local contribution 
\begin{equation}
\begin{aligned}
    \hbar\Sigma^{\star}_\text{nl}=& \frac{3g_d}{2}\left[\cos^2\theta\left(\nabla^2_\mathbf{r}\right)-\sin^2\theta\frac{\partial^2}{\partial r_y^2}\right]U_\text{2D}(\mathbf{r})*_\mathbf{r}n_0(\mathbf{r})\\
    &+f_0(\mathbf{r},\mathbf{k})*_\mathbf{k} U^{\text{nl}}_\text{eff}(\mathbf{k})/(2\pi)^2,
\label{Sigma}
\end{aligned}
\end{equation}
where $*_{\mathbf{r}/\mathbf{k}}$ signifies the convolution operation on the real-space/momentum-space argument, and $U_\text{2D}(\mathbf{r})={e^{r^2/4l_z^2}}K_0(r^2/4l_z^2)/{(2\pi)^{3/2}}$ (see End Matter and Ref.~\cite{cai2010mean}) with $K_0$ the modified Bessel function of the second kind. 
The total areal density profile that can be directly compared with the \textit{in-situ} absorption images is
\begin{equation}
n(\mathbf r)=\sum_{j=0}^{\infty} n_{j}(\mathbf r),\quad
n_{j}(\mathbf r)=\int\!\frac{d^{2}k}{(2\pi)^{2}}\,f_{j}(\mathbf r,\mathbf k),
\end{equation}
while only \(n_{0}(\mathbf r)\) feeds back into the mean field via Eqs.~(\ref{phase_space_distribution}) and (\ref{Sigma}).
The chemical potential $\mu$ is self-consistently determined by $N=\int d^2r n(\mathbf r,\mu)$. 

If the von-Weizs\"acker term \(f^{\mathrm{vW}}\) is neglected, our theory reduces to a quasi-2D generalization of the Hartree-Local-Fock theory~\cite{zhang2010thermodynamic,baillie2012local}, i.e., the standard LDA for contact-interacting gases~\cite{tung2010observation,hung2011observation,yefsah2011exploring,plisson2011coherence,fletcher2015connecting} augmented by the nonlocal component of the proper self-energy, \(\Sigma^{\star}_\text{nl}\).
In this limit, since the anisotropic nature of the Fock contribution [second term in Eq.~(\ref{Sigma})] is exactly integrated out, any real-space anisotropy can only arise from the Hartree part of \(\Sigma^{\star}_\text{nl}\) [first term in Eq.~(\ref{Sigma})].
For quasi-2D clouds with slowly varying density, the Hartree term is small due to the derivative form in Eq.~(\ref{Sigma}), and vanishes in the homogeneous limit.
As a result, the overall magnetostriction of the thermal profile is negligible at the Thomas-Fermi level.

Beyond the Thomas-Fermi level, anisotropic deformation can also enter through the von-Weizs\"acker quantum correction \(f^{\mathrm{vW}}\)~\cite{bartel1985extended,vanzyl2013thomasfermivon}.
However, the interaction-induced contribution to \(f^{\mathrm{vW}}\) is parametrically suppressed by \((\lambda_T/l_r)^4(a_{dd}/\lambda_T)^2\!\ll\!1\) (see End Matter), where \(\lambda_T=\sqrt{2\pi\hbar^2/mk_BT}\) is the thermal wavelength and \(l_r=\sqrt{\hbar/m(\omega_x\omega_y)^{1/2}}\) is the radial oscillator length.
Higher-order terms in the gradient expansion are even more strongly suppressed under experimentally relevant finite-temperature conditions.
Consequently, magnetostriction in the normal phase is practically negligible even within the full HFMF treatment, in agreement with our observations [Fig.~\ref{insitu}(b)].

In contrast, at the Thomas-Fermi level of the dipolar GPE, the density profile obeys
\begin{equation}
    \left[ V(\mathbf{r})+\int \frac{d^2k}{(2\pi)^2}\,e^{-i\mathbf{k}\cdot\mathbf{r}}\,U_{\text{eff}}(\mathbf{k})\,\rho(\mathbf{k})-\mu\right]\phi(\mathbf{r})=0,
\end{equation}
where \(\phi(\mathbf{r})=\sqrt{N_0}\Psi(\mathbf{r})\) is the macroscopic wavefunction with \(N_0\) the total number of particles in the superfluid, and \(\rho(\mathbf{k})=\int d^2r\,e^{i\mathbf{k}\cdot\mathbf{r}}\,|\phi(\mathbf{r})|^2\).
Because the superfluid possesses global coherence, the order-parameter profile directly samples the anisotropy of \(U_{\text{eff}}(\mathbf{k})\) via this convolution in momentum space, yielding pronounced magnetostriction.

\begin{figure}[t!]
\includegraphics[width=1\columnwidth]{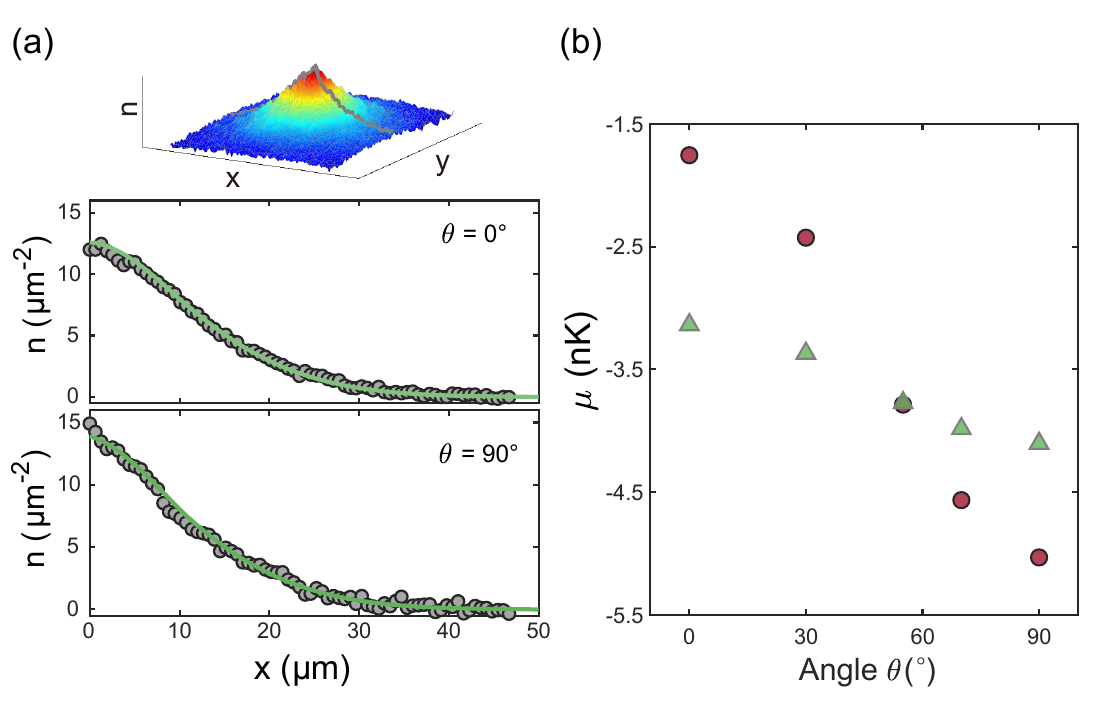}% Here is how to import EPS art
\caption{\textbf{Thermometry across dipole angles.} (a) Vertical cut of \textit{in-situ} density profile of normal gas at $\theta=0^\circ$ and $\theta=90^\circ$~(gray circles) and the fitted profiles~(green curves). (b) Green triangles are the fitted chemical potential as a function of $\theta$, including the nonlocal term; red circles are the chemical potential calculated with only local $g_{\text{eff}}$. }
\label{profile} 
\end{figure}

\textit{Normal Gas Thermometry.---} The magnetostriction-free character of the normal phase enables reliable thermometry. Varying $(N,T)$,
we fit the experimental profiles using the HFMF formalism evaluated with fast Fourier transforms. In Fig.~\ref{profile}(a), for the same atom number $N\simeq 12200$ and temperature $T\simeq 70.2~\text{nK}$, the FFT-HFMF fits accurately reproduce the full set of normal-gas density profiles measured at different dipole angles $\theta$.
Green triangles in Fig.~\ref{profile}(b) show the chemical potentials extracted from the fits as a function of $\theta$. To illustrate the contribution from the nonlocal term, we also show results with $\hbar\Sigma^\star_\text{nl}$ artificially set to zero (red circles). We observe that when $\theta$ is smaller~(larger) than $55^\circ$, the nonlocal interaction contributes an additional negative (positive) chemical potential, corresponding to attractive (repulsive) effects. This yields a mild systematic correction to contact-based thermometry that considers only $g_{\text{eff}}$. 

\begin{figure}[t!]
\includegraphics[width=0.95\columnwidth]{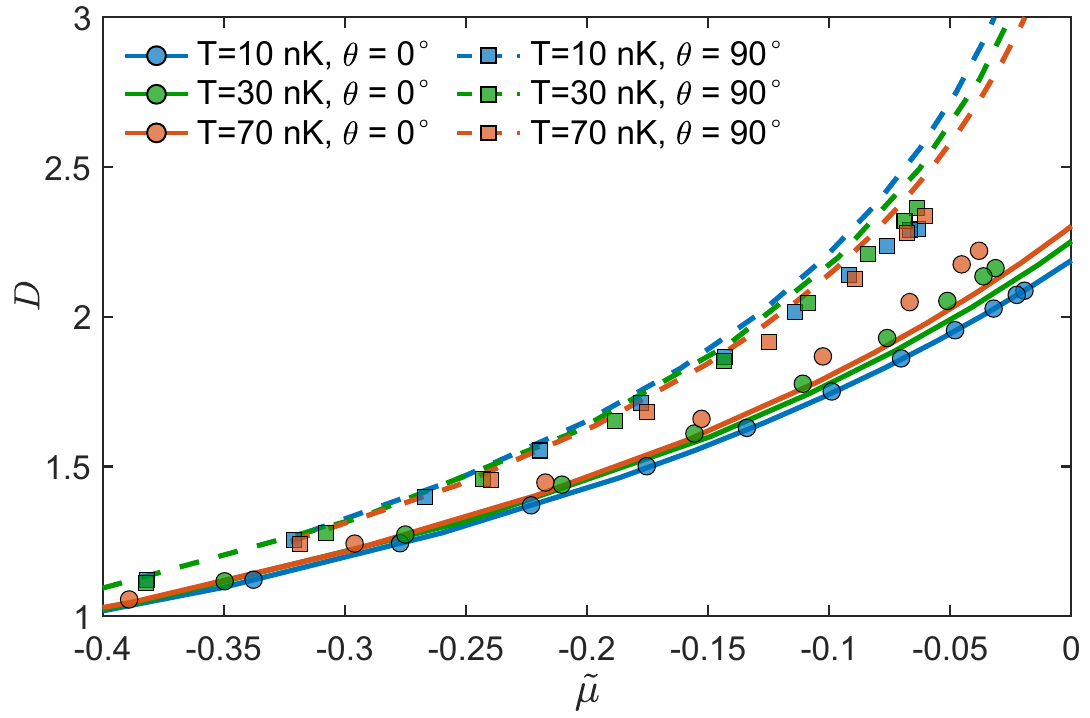}% Here is how to import EPS art
\caption{\textbf{EOS and scale-invariance breaking.} Numerically evaluated EOS including nonlocal interaction. Lines denote results directly evaluated using homogeneous systems, discrepancies among different temperatures show the breaking of scale invariance. Symbols are extracted results from trapped systems based on LDA, with $T/T_c \simeq 1.1$ and other parameters identical to those in our experimental setup. }
\label{EOS} 
\end{figure}

\textit{Equation of State.---}
The magnetostriction-free thermal wings also allow an approximately LDA-based extraction of the EOS, whose validity in the presence of nonlocal interactions must be checked explicitly.
For a homogeneous 2D Bose gas with contact interactions, the EOS possesses scale invariance~\cite{pitaevskii2016bose}: introducing the phase-space density $D\equiv n_{\mathrm{hom}}\lambda_{T}^{2}$, the normal-state EOS reads
\begin{equation}
D=-\ln\!\left[1-\exp(\tilde{\mu}-\tilde{g}D/\pi)\right],
\label{EOS_contact}
\end{equation}
where $\tilde{\mu}\equiv\mu/k_{B}T$ and $\tilde{g}=m g/\hbar^{2}$. Under the LDA with $D(\mathbf{r})=n_{0}(\mathbf{r})\lambda_{T}^{2}$ and $\mu(\mathbf{r})=\mu-V(\mathbf{r})$, the EOS for purely local coupling coincides with Eq.~(\ref{EOS_contact}), but for nonlocal interactions this need not hold since the density depends on a nonlocal self-energy. As shown in Fig.~\ref{EOS}, numerically evaluating the homogeneous EOS at different temperatures yields curves that do not collapse, evidencing broken scale invariance. However, trapped-system extractions converge to the homogeneous EOS in the low-density wings (sufficiently negative $\tilde{\mu}$), validating the practical use of LDA in this dilute regime.

\textit{Magnetostriction Crossover near Superfluid Transition.---}
\begin{figure}[t!]
\includegraphics[width=1\columnwidth]{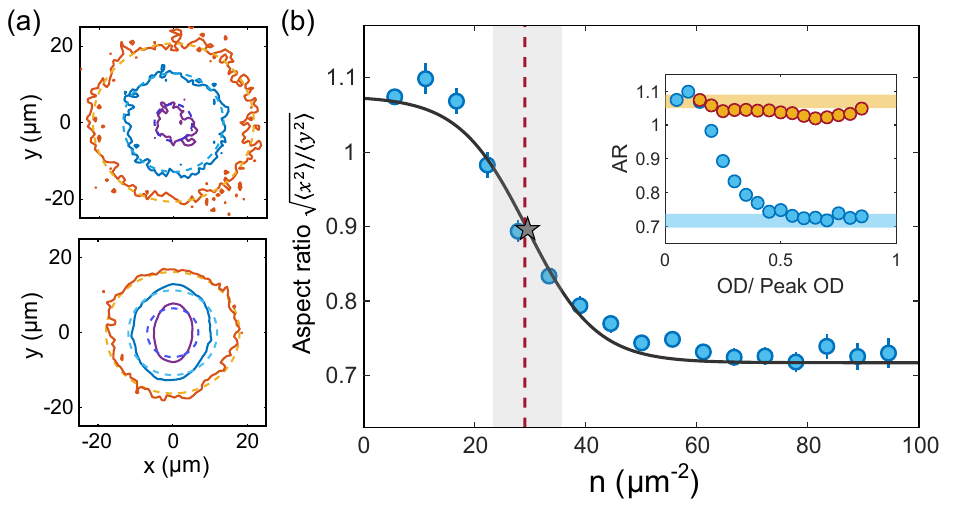}% Here is how to import EPS art
\caption{\textbf{Observation of magnetostriction crossover near the superfluid transition.} (a) Contours of atomic density at $\theta=90^\circ$~(solid lines) and harmonic potential~(dashed lines) of normal gas~(top) and normal-superfluid mixtures~(bottom). (b) Aspect ratio of the superfluid sample versus areal density. The solid curve is a phenomenological sigmoid fit with crossover center $n_0=29.5~\mu\text{m}^{-2}$~(pentagram) and width $w=6.2~\mu\text{m}^{-2}$~(gray shaded region). The dashed line indicates the na\"ive BKT critical density estimate $n_c\simeq29.0~\mu\text{m}^{-2}$. Error bars are standard errors from 25 measurements. Inset:  Aspect ratio of contours for different densities. Orange circles are for purely normal gas, and blue circles are for superfluid gas. Shaded orange~(blue) regime represents the aspect ratio of trap~(Thomas-Fermi radius) with fitting errors.}
\label{AR}
\end{figure}
Having established that the normal phase is magnetostriction-free and that LDA-based thermometry remains valid in our sample, we now turn to the crossover between the isotropic thermal cloud and the magnetostrictive superfluid. We image density contours near the superfluid transition at $\theta=90^\circ$, as shown in Fig.~\ref{AR}(a) for both a purely normal gas and a normal-superfluid mixture. We quantify spatial anisotropy by the aspect ratio $\sqrt{\langle x^2\rangle/\langle y^2\rangle}$ of each contour [inset of Fig.~\ref{AR}(b)]. The purely thermal gas exhibits minimal anisotropy: the aspect ratio of its contours closely matches that of the trap, confirming its magnetostriction-free character. In contrast, for the superfluid core embedded in a thermal background, we observe a smooth crossover of aspect ratio from the trap limit to the Thomas-Fermi limit. Away from the trap center, the density distribution follows equipotential curves. As the density increases, the aspect ratio gradually decreases. Phenomenologically, we capture this crossover by a sigmoid fit $\text{AR}(n)=\text{AR}_\text{TF}+(\text{AR}_\text{trap}-\text{AR}_\text{TF})/(1+e^{(n-n_0)/w})$ to the single-shot-averaged aspect ratio versus density [Fig.~\ref{AR}(b)], where the aspect ratio $\sqrt{\langle x^2\rangle/\langle y^2\rangle}$ of the region above each density threshold is computed per shot and averaged over 25 measurements. The fit yields a crossover center $n_0=29.5~\mu\text{m}^{-2}$ and width $w=6.2(7)~\mu\text{m}^{-2}$, with $\text{AR}_\text{trap}=1.065$ and $\text{AR}_\text{TF}=0.717$. This sigmoid form admits a heuristic two-fluid interpretation: defining $f_{qc}(n)=1/[1+e^{-(n-n_0)/w}]$, we write the observed aspect ratio as $\text{AR}=f_{qc}\,\text{AR}_\text{TF}+(1-f_{qc})\,\text{AR}_\text{trap}$, where $f_{qc}$ plays the role of an effective quasi-condensate fraction that interpolates between the isotropic normal phase and the anisotropic Thomas-Fermi limit. Notably, $n_0$ is close to a na\"ive estimate of the Berezinskii-Kosterlitz-Thouless (BKT) critical density $n_c=\mathcal{D}_c/\lambda_T^2\simeq29.0~\mu\text{m}^{-2}$, obtained from $\mathcal{D}_c=\ln(380/\tilde{g})$~\cite{prokof2001critical} using the local coupling $\tilde{g}=mg_\text{eff}/\hbar^2\simeq0.0016$ at $\theta=90^\circ$ and $T\simeq43~\text{nK}$~\cite{he2025exploring}.  The proximity of $n_0$ and $n_c$ suggests that the crossover is tied to the buildup of coherence enabling magnetostriction. Beyond the transition, when phase coherence is fully established, the system follows the Thomas-Fermi description of the dipolar GPE, and strong DDI-induced anisotropy manifests. This density-dependent crossover provides a direct observable for tracking how anisotropic interactions reshape the cloud as coherence develops, opening a window into beyond-mean-field physics near the 2D superfluid transition.

\textit{Conclusion.---} To summarize, we have examined density profiles of strongly dipolar quasi-2D gases in harmonic traps in both purely normal and normal-superfluid regimes. We reveal that thermal clouds with temperature close to quantum degeneracy remain magnetostriction-free even for strongly tilted dipoles, contrasting sharply with the robust magnetostriction of the superfluid core. This behavior is attributed to the inability of the nonlocal Fock term to generate spatial anisotropy at the Thomas-Fermi level in the normal phase, the smallness of the nonlocal Hartree contribution due to its derivative form for slowly varying density profiles, and the thermal suppression of von-Weizs\"acker gradient corrections. We establish a thermometry method based on HFMF theory, including both local and nonlocal DDI contributions, which fits normal-phase density profiles across different dipole angles with a single $(T,N)$ pair. We find that LDA-based EOS extraction from trapped systems remains valid for the thermal wing of the strongly dipolar gases in quasi-2D.
This approach is, in principle, applicable to other quantum states such as 2D quantum droplets~\cite{wenzel2017striped} or supersolids~\cite{he2025observation} for studies of their thermodynamics. The observed aspect ratio crossover from the magnetostriction-free periphery to the anisotropic core indicates that beyond-mean-field fluctuations~\cite{PhysRevA.86.053602,prokof2002two,lima2011quantum} begin imprinting anisotropy before full superfluid coherence. Tracking aspect ratio across density contours may enable future precision studies of how nonlocal anisotropic interactions affect BKT physics.

\begin{acknowledgments}
\paragraph*{\bf Acknowledgement}
This work is supported by Hong Kong RGC (No. RFS2122-6S04 and C4050-23GF). GBJ acknowledges support from
the Gordon and Betty Moore Foundation, grant DOI
10.37807/GBMF13794. YY and XYG acknowledge support from the National Natural Science Foundation of China under Grant No. 124B2074, 92565105, 
Hong Kong RGC No. 14301425, and No. 24308323, 
Guangdong Provincial Quantum Science Strategic Initiative GDZX2404004, GDZX2505005, and 
the Space Application System of China Manned Space Program. 
\end{acknowledgments}

\section*{DATA AVAILABILITY}
The data that support the findings of this article are openly available at \cite{data}. 

\newpage
\section{End Matter}
\phantomsection
The von-Weizs\"acker term in Eq.~(\ref{phase_space_distribution}) is given by
\begin{equation}
    f_j^{\text{vW}}=-\frac{\{\{{E}_j,{E}_j\}\}}{16} \frac{\partial^2 g_{B}({E}_j)}{\partial {{E}_j}^2}-\frac{[[{E}_j,{E}_j]]}{24}\frac{\partial^3 g_{B}({E}_j)}{\partial {{E}_j}^3}.
\end{equation}
where supposing $A(\mathbf{r},\mathbf{k})$ and $B(\mathbf{r},\mathbf{k})$ are two phase-space quantities, the double Poisson bracket of $A$ and $B$ is defined as
\begin{align}
    \{\{{A},&{B}\}\}
    =
    \sum_{i,j\in\{x,y\}}
    \Bigl(
        \partial_{r_i}\partial_{r_j}{A}\;
        \partial_{k_i}\partial_{k_j}{B} \nonumber\\
        &+ \partial_{k_i}\partial_{k_j}{A}\;
        \partial_{r_i}\partial_{r_j}{B}
        -2\partial_{r_i}\partial_{k_j}{A}\;
        \partial_{k_i}\partial_{r_j}{B}
    \Bigr),
\label{eq:double_poisson}
\end{align}
and the auxiliary bracket is
\begin{align}
    [[{A},&{B}]]
    =\sum_{i,j\in\{x,y\}}
    \Bigl(
    \partial_{r_i}\partial_{r_j}A\;
    \partial_{k_i}B\partial_{k_j}B \nonumber \\
    & + \partial_{k_i}\partial_{k_j}A\;
    \partial_{r_i}B\partial_{r_j}B
    -2\partial_{r_i}\partial_{k_j}A\;
    \partial_{k_i}B\partial_{r_j}B
\Bigr).
\label{eq:auxiliary_bracket}
\end{align}
Though anisotropic deformation can also enter through the von-Weizs\"acker quantum correction \(f^{\mathrm{vW}}\) because the double-bracket operations couple momentum and real space, thereby transferring the strong momentum-space anisotropy generated by the Fock term into the real-space profile. Below, we show that this term is suppressed by a small prefactor under our experimental conditions by providing detailed derivation of the theoretical framework adopted for the normal (thermal) gas in the main text.

The lesser Green's function is defined as
\begin{equation}
    G^{<}_{j j'}(\mathbf{r},t,\mathbf{r}',t')=-i\langle\hat\psi_j^{\dagger}(\mathbf{r}',t')\hat\psi_{j'}(\mathbf{r},t)\rangle,
\end{equation}
where $\hat \psi_j$ is the bosonic field operator in the axial $j$th band, which obeys canonical bosonic commutation relations. In frequency space, we denote
\begin{equation}
    G^{<}_{j j'}(\mathbf{r},\mathbf{r}',\omega)=i\int dt e^{i \omega t} G_{j j'}^{<}(\mathbf{r},\mathbf{r}', 0 ).
\end{equation}
The lesser Green's function is directly related to the one-body density matrix via
\begin{equation}
\begin{split}
    \langle n_{jj'}(\mathbf{r},\mathbf{r}')\rangle&=\int\frac{d\omega}{2\pi}G^{<}_{jj'}(\mathbf{r},\mathbf{r}',\omega)\\
    &=\int\frac{d\omega}{2\pi}g_B(\hbar\omega) A_{jj'}(\mathbf{r},\mathbf{r}',\omega),
\end{split}
    \label{SM:one-body_density_matrix_and_A}
\end{equation}
where $A_{jj'}$ is the spectral function, which can be obtained from the retarded Green's function $G^R$,
\begin{equation}
    G^R_{jj'}(\mathbf{r},t,\mathbf{r}',t')=-i\langle[\hat\psi_j(\mathbf{r},t),\hat\psi_{j'}^{\dagger}(\mathbf{r}',t')]\rangle\theta(t-t').
\end{equation}
The Dyson equation for $G^R$ in mixed (frequency\textendash real) space can be written in terms of the retarded proper self-energy $\Sigma^{\mathrm{R} \star}$~\cite{fetter2012quantum}
\begin{multline}
    [\omega-\hbar^{-1}\varepsilon^{(0)}_{j,\mathbf k}-\hbar^{-1}V(\mathbf{r})+i\epsilon]G^{\mathrm R}_{jj'}(\mathbf{r},\mathbf{r}',\omega)\\
    =\delta(\mathbf{r}-\mathbf{r}')+\int d^2x \Sigma^{\mathrm{R} \star}_{jj'}(\mathbf{r},\mathbf{x})G^{\mathrm R}_{jj'}(\mathbf{x},\mathbf{r}',\omega)
    \label{SM:dyson_eqn}
\end{multline}
where $\epsilon$ is a positive infinitesimal and $\Sigma^{\mathrm{R} \star}_{jj'}(\mathbf{r},\mathbf{r}')$ is the retarded self-energy. Under the Hartree\textendash Fock approximation, we retain only the effective interaction $U_\text{eff}$ within the $j=0$ band, and the self-energy reads~\cite{fetter2012quantum}
\begin{multline}
    \Sigma^{\mathrm{R} \star}_{jj'}(\mathbf{r},\mathbf{r}')=\delta(\mathbf{r}-\mathbf{r}')\delta_{jj'}\int d^2x U_\text{eff}(\mathbf{r}-\mathbf{x})\langle n_j(\mathbf{x})\rangle\\
    +\delta_{jj'}U_\text{eff}(\mathbf{r}-\mathbf{r}')\langle n_{jj'}(\mathbf{r},\mathbf{r}')\rangle
    \label{SM:HF_self_energy_real}
\end{multline}
where $\langle n_j(\mathbf{x})\rangle$ denotes the number density at position $\mathbf{x}$ in band $j$. For simplicity, we henceforth write $\Sigma^{\star}\equiv \Sigma^{\mathrm{R} \star}_{00}$, $G^R_j\equiv G^R_{jj}$ and $A_j\equiv A_{jj}$, since within Hartree\textendash Fock all bands share the same self-energy. The Wigner transformation of a quantity $O(\mathbf{r},\mathbf{r}')$ is defined as
\begin{equation}
    {O}(\mathbf{r}, \mathbf{k})=\int d^2 s e^{-i \mathbf{k} \cdot \mathbf{s} } O(\mathbf{r}+\mathbf{s}/2,\mathbf{r}-\mathbf{s}/2).
\end{equation}
After the Wigner transformation, Eqs.~(\ref{SM:dyson_eqn}) and (\ref{SM:HF_self_energy_real}) become

\begin{gather} \begin{split} &\Bigl[\, \omega-\hbar^{-1}\varepsilon^{(0)}_{j,\mathbf k} -\hbar^{-1}V(\mathbf r)\\ &\quad -\delta_{j,0}{\Sigma}^{\star}(\mathbf r,\mathbf k,\omega)+i\epsilon \Bigr]\!\star\! {G}_j^{\mathrm R}=1, \end{split} \label{SM:transformed_dyson_eqn} \\ \begin{aligned} {\Sigma}^{\star}(\mathbf r,\mathbf k,\omega)=&\,2g_\text{eff} \int\frac{d^2q}{(2\pi)^2} f_{0}(\mathbf{r},\mathbf{q})\\ &+ \int d^2x U^\text{nl}_\text{eff}(\mathbf{r}-\mathbf{x})n_0(\mathbf{x}) \\ &+ \int\frac{d^2q}{(2\pi)^2} f_{0}(\mathbf{r},\mathbf{q}) U_\text{eff}^\text{nl}(\mathbf{k}-\mathbf{q})\\ =&\,2g_\text{eff}n_0(\mathbf{r})+U_\text{eff}^\text{nl}(\mathbf{r})(*)_\mathbf{r} n_0(\mathbf{r})\\ &+f_{j}(\mathbf{r},\mathbf{k})(*)_\mathbf{k} U_\text{eff}^\text{nl}(\mathbf{k})/(2\pi)^2, \end{aligned} \label{SM:HF_self_energy} \end{gather}

Because $U_\text{eff}^\text{nl}(\mathbf{r})$ diverges at $\mathbf{r}=0$, a convenient regularized form for the real-space convolution is~\cite{cai2010mean}
\begin{multline}
    U_\text{eff}^\text{nl}(\mathbf{r})(*)_\mathbf{r} n_0(\mathbf{r})\equiv -\frac{3g_d}{2}\biggl[\sin^2\theta\frac{\partial^2}{\partial r_y^2}\\
    -\cos^2\theta\!\left(\frac{\partial^2}{\partial r_y^2}+\frac{\partial^2}{\partial r_x^2}\right)\biggr]U_\text{2D}(\mathbf{r})(*)_\mathbf{r}n_0(\mathbf{r}),
\end{multline}
We define the $j$th-band phase-space density $f_j(\mathbf{r},\mathbf{k})$ as the Wigner transform of the one-body density matrix,
\begin{equation}
    f_j(\mathbf{r}, \mathbf{k})=\int d^2 s e^{-i \mathbf{k} \cdot \mathbf{s} } \langle n_j(\mathbf{r}+\mathbf{s}/2,\mathbf{r}-\mathbf{s}/2)\rangle.
\end{equation}
The operation $\star$ in Eq.~(\ref{SM:transformed_dyson_eqn}) is the Moyal product, which admits a gradient expansion. Writing phase-space functions with momentum $\mathbf{p}=\hbar\mathbf{k}$ as $\tilde{A}(\mathbf{r},\mathbf{p})\equiv A(\mathbf{r},\mathbf{k})$ and similarly for $\tilde{B}$,
\begin{equation}
    A\star B\equiv \tilde{A}\star\tilde{B}=\tilde{A}\tilde{B}+\frac{i\hbar}{2}\{\tilde{A},\tilde{B}\}_c   -\frac{\hbar^2}{8}\{\!\{\tilde{A},\tilde{B}\}\!\}_c+\mathcal{O}(\hbar^3),
\end{equation}
where $\{\tilde{A},\tilde{B}\}_c\equiv 
{\partial}_{\mathbf{r}}\tilde{A}\!\cdot\!{\partial}_{\mathbf{p}}\tilde{B}-
{\partial}_{\mathbf{p}}\tilde{A}\!\cdot\!{\partial}_{\mathbf{r}}\tilde{B}$ is the classical Poisson bracket, and $\{\!\{\cdot,\cdot\}\!\}_c$ has the same form as Eq.~(\ref{eq:double_poisson}) with $\mathbf{k}\to\mathbf{p}$.
We expand the retarded Green function in powers of $\hbar$,
\begin{equation}
\tilde{G}^{\mathrm R}_j=\tilde{G}^{(0)}_j+\hbar \tilde{G}^{(1)}_j+\hbar^{2}\tilde{G}^{(2)}_j+\mathcal{O}(\hbar^{3}).
\end{equation}
For convenience, we denote
\begin{equation}
\begin{split}
    &\tilde{M}_j(\mathbf{r},\mathbf{p},\omega)\equiv \omega - \tilde{E}_j(\mathbf{r},\mathbf{p})+i\epsilon\\
    &=\omega-\hbar^{-1}\varepsilon^{(0)}_{j,\mathbf p}
        -\hbar^{-1}V(\mathbf r)-\delta_{j,0}\tilde{\Sigma}^{\star}(\mathbf r,\mathbf p)+i\epsilon.
\end{split}
\end{equation}
Keeping only the zeroth-order ($\tilde{A}\tilde{B}$) term of the Dyson equation gives
$\tilde{M}_j\,\tilde{G}^{(0)}_j=1$, hence
\begin{equation}
\tilde{G}^{(0)}_j={1}/{\tilde{M}_j}.
\end{equation}
Collecting the $\mathcal{O}(\hbar)$ terms, we obtain  
$\tilde{M}_j\,\tilde{G}^{(1)}_j+\dfrac{i}{2}\{\tilde{M}_j,\tilde{G}^{(0)}_j\}=0$, and since $\{\tilde{M}_j,\tilde{G}^{(0)}_j\}_c=0$, i.e.
\begin{equation}
\tilde{G}^{(1)}_j=-\frac{i}{2}\,\frac{1}{\tilde{M}_j}\{\tilde{M}_j,\tilde{G}^{(0)}_j\}_c=0.
\end{equation}
At $\mathcal{O}(\hbar^{2})$, we have
\begin{gather}
\tilde{M}_j\,\tilde{G}^{(2)}_j-\tfrac18\{\!\{\tilde{M}_j,\tilde{G}^{(0)}_j\}\!\}_c=0\nonumber\\
\Longrightarrow\quad
\tilde{G}^{(2)}_j=\frac1{8\tilde{M}_j}\,\{\!\{\tilde{M}_j,\tilde{G}^{(0)}_j\}\!\}_c.
\end{gather}
Using $G^{(0)}=1/\tilde{M}_j$ we require $\{\!\{\tilde{M}_j,1/\tilde{M}_j\}\!\}_c$:
\begin{equation}
\begin{split}
&\{\{\tilde{M}_j,1/\tilde{M}_j\}\}_c=-\{\{\ln\tilde{M}_j,\ln\tilde{M}_j\}\}_c\\
&\quad=2\frac{[[\tilde{M}_j,\tilde{M}_j]]_c}{\tilde{M}_j^3}-\frac{\{\{\tilde{M}_j,\tilde{M}_j\}\}_c}{\tilde{M}_j^2},
\end{split}
\end{equation}
where $[[\cdot,\cdot]]_c$ is defined analogously to Eq.~(\ref{eq:auxiliary_bracket}) with $\mathbf{k}\to\mathbf{p}$.
Noting that
\begin{equation}
\begin{aligned}
    \{\{\tilde{M}_j,\tilde{M}_j\}\}_c&=\{\{-\tilde{E}_j,-\tilde{E}_j\}\}_c=\{\{\tilde{E}_j,\tilde{E}_j\}\}_c,\\
    [[\tilde{M}_j,\tilde{M}_j]]_c&=[[-\tilde{E}_j,-\tilde{E}_j]]_c=-[[\tilde{E}_j,\tilde{E}_j]]_c,
\end{aligned}
\end{equation}
\begin{equation}
\tilde{G}^{(2)}_j=
-\frac{1}{4}\frac{[[\tilde{E}_j,\tilde{E}_j]]_c}{\tilde{M}_j^{4}}-\frac{1}{8}\frac{\{\!\{\tilde{E}_j,\tilde{E}_j\}\!\}_c}{\tilde{M}_j^{3}}.
\end{equation}
Consequently,
\begin{equation}
\begin{split}
\tilde{G}^{\mathrm R}_j=&
\frac{1}{\omega-\tilde{E}_j+i\epsilon}
-
\frac{\hbar^2}{8}
\frac{\{\{\tilde{E}_j,\tilde{E}_j\}\}_c}{(\omega-\tilde{E}_j+i\epsilon)^{3}}\\
&-
\frac{\hbar^2}{4}\frac{[[\tilde{E}_j,\tilde{E}_j]]_c}{(\omega-\tilde{E}_j+i\epsilon)^{4}}
+\mathcal{O}(\hbar^4),
\end{split}
\end{equation}
Absorbing factors of $\hbar^2$ into the double-bracket operations, $\{\{\cdot,\cdot\}\}=\hbar^2\{\{\cdot,\cdot\}\}_c$ and $[[\cdot,\cdot]]=\hbar^2[[\cdot,\cdot]]_c$, yields
\begin{equation}
\begin{split}
    G^{\mathrm R}_j=&\;
\frac{1}
     {\omega-E_j+i\epsilon}
-
\frac{1}{8}
\frac{
   \{\!\{E_j,\;E_j\}\!\}
     }
     {(\omega-E_j+i\epsilon)^{3}}\\
&-
\frac{1}{4}
\frac{
   [[E_j,\;E_j]]
     }
     {(\omega-E_j+i\epsilon)^{4}}
+\cdots .
\end{split}
\end{equation}
The spectral function obeys $A_j=-2\,\mathrm{Im}\,G^{\mathrm R}_j$, thus
\begin{equation}
\begin{split}
    A_j(\mathbf R,\mathbf k,\omega)=&\;
    2\pi\,\delta(\omega\!-\!E_j)\\
    &-\frac{\pi}{8}\{\!\{E_j,E_j\}\!\}
    \partial_{\omega}^2\delta(\omega\!-\!E_j)\\
    &+\frac{\pi}{12}[[E_j,E_j]]\partial_\omega^{3}\delta(\omega\!-\!E_j)+\cdots.
\end{split}
\end{equation}
From Eq.~(\ref{SM:one-body_density_matrix_and_A}) we obtain the relation between $f_j$ and $A_j$
\begin{equation}
    f_j(\mathbf r,\mathbf k)=
\!\int\!\frac{d\omega}{2\pi}\,g_{B}(\hbar\omega)\,A_j(\mathbf r,\mathbf k,\omega).
\end{equation}
Evaluating $\int d\omega\,\partial_{\omega}^{n}\delta(\hbar\omega-E)\,g_{B}(\hbar\omega)=(-\partial_{E})^{n}g_{B}(E)$ gives
\begin{equation}
f_j=f^{\text{TF}}_j+f^{\text{vW}}_j+\mathcal{O}(\hbar^{4}),
\end{equation}
which reproduces Eq.~(\ref{phase_space_distribution}) in the main text. When the nonlocal contribution to the proper self-energy is neglected, i.e., $\Sigma^\star(\mathbf{r},\mathbf{k})=\Sigma^\star_\text{l}(\mathbf{r})=2g_{\text{eff}}n_0(\mathbf{r})$, the Thomas\textendash Fermi phase-space distribution $f^{\text{TF}}_j(\mathbf{r},\mathbf{k};\Sigma^\star_\text{nl}=0)$ reduces to the usual LDA form for contact gases:
\begin{multline}
    \biggl[\exp\!\biggl(\frac{\hbar^2k^2/2m+j\hbar\omega_z-\mu(\mathbf{r})+2g_{\text{eff}}n_0}{k_B T}\biggr)\!-\!1\biggr]^{\!-1}\!,
\end{multline}
where $\mu(\mathbf{r})\equiv\mu-V(\mathbf{r})$. At the TF level, the self-consistent solution $n(\mathbf{r})$ depends only on the local chemical potential $\mu(\mathbf{r})$, so the profile follows the trap.

For numerical implementation, we adopt dimensionless variables based on the thermal wavelength: lengths in units of $\lambda_T$ and momenta in units of $1/\lambda_T$. Defining $\bar{\mathbf{r}}=\mathbf{r}\lambda_T/l_r^2$ and $\bar{\mathbf{k}}=\mathbf{k}\lambda_T$, we have
\begin{align}
    &f_j^{\text{LDA}}(\bar{\mathbf{r}},\bar{\mathbf{k}})=\frac{1}{e^{{\mathcal{E}_j}}z^{-1}-1}, \\
    f^{\text{vW}}_j=&-\frac{1}{16}\frac{\lambda_T^4}{l_r^4}\{\{{\mathcal{E}_j},{\mathcal{E}_j}\}\}_T\biggl[\frac{2e^{2{\mathcal{E}_j}}z^{-2}}{(e^{{\mathcal{E}_j}}z^{-1}\!-\!1)^3}-\frac{e^{{\mathcal{E}_j}}z^{-1}}{(e^{{\mathcal{E}_j}}z^{-1}\!-\!1)^2}\biggr] \nonumber \\
    &-\frac{1}{24}\frac{\lambda_T^4}{l_r^4}[[{\mathcal{E}_j},{\mathcal{E}_j}]]_T\biggl[\frac{6e^{2{\mathcal{E}_j}}z^{-2}}{(e^{\mathcal{E}_j}z^{-1}\!-\!1)^3} \nonumber\\
\end{align}
where the dimensionless single-particle energy is $\mathcal{E}_j=(\bar{r}^2+\bar{k}^2)/4\pi+\delta_{j,0}\hbar \Sigma^\star/k_B T-{j \lambda_T^2}/{2\pi l_z^2}$. The fugacity is $z=e^{\mu/k_BT}$. We write $\{\{\cdot,\cdot\}\}_T=(l_r/\lambda_T)^4\{\{\cdot,\cdot\}\}$ and $[[\cdot,\cdot]]_T=(l_r/\lambda_T)^4[[\cdot,\cdot]]$ in terms of the dimensionless variables $\bar{\mathbf{r}}$ and $\bar{\mathbf{k}}$. Since $\hbar \Sigma^\star/k_B T\propto g/(\lambda_T^3 k_B T)$, the interaction correction to $f_j$ scales as $(\lambda_T^4/l_r^4)(g/(\lambda_T^3 k_B T))^2\propto (\lambda_T^4/l_r^4)(a_{dd}/\lambda_T)^2$.

\newpage
%\bibliography{references}% Produces the bibliography via BibTeX.
\providecommand{\noopsort}[1]{}\providecommand{\singleletter}[1]{#1}%

\end{document}